\documentclass[11pt ]{article}
\usepackage[affil-it]{authblk}

\usepackage{graphicx}
\usepackage{amsmath, amsthm, amssymb}
\usepackage{subfigure}
\usepackage{comment}
\usepackage{bm}
\usepackage{epsfig,color}
\usepackage[sans]{dsfont}
\usepackage{verbatim}
\usepackage{amsfonts}
\usepackage{amsmath}
\usepackage{amssymb}
\usepackage{enumerate}

\newcommand{\be}{\begin{equation}}
\newcommand{\ee}{\end{equation}}
\newcommand{\bea}{\begin{eqnarray}}
\newcommand{\eea}{\end{eqnarray}}
\newcommand{\bes}{\begin{equation*}}
\newcommand{\ees}{\end{equation*}}
\newcommand{\beas}{\begin{eqnarray*}}
\newcommand{\eeas}{\end{eqnarray*}}

\newcommand{\x}{\mathrm{x}}

\def\x{\mathrm{x}}
\def\y{\mathrm{y}}
\def\w{\mathrm{w}}
\def\g{\mathrm{guess}}

\newtheorem*{thm*}{Theorem}

\newtheorem*{lem*}{Lemma}

\newtheorem*{lipschitzLem*}{Lemma \ref{lipschitz}}
\newtheorem*{lipschitzCubeLem*}{Lemma \ref{lipschitzCube}}
\newtheorem*{pgmNearlyOptimalThm*}{Theorem \ref{pgmNearlyOptimal}}
\begin{document}

\title{Structure of Optimal State Discrimination in Generalized Probabilistic Theories}

\author{Joonwoo Bae$^{1,2}\footnote{bae.joonwoo@gmail.com}$, D.-G. Kim$^{2}$ and Leong-Chuan Kwek$^{3,4,5,}$}

\affil{
$^{1}$ Department of Applied Mathematics, Hanyang University (ERICA), 55 Hanyangdaehak-ro, Ansan, Gyeonggi-do, 426-791, Korea\\
$^{2}$ Freiburg Institute for Advanced Studies (FRIAS), Albert-Ludwigs University of Freiburg, Albertstrasse 19, 79104 Freiburg, Germany \\
$^{3}$ Center for Quantum Technologies, National University of Singapore, \\ 3 Science Drive 2,  Singapore 117543 \\
$^{4}$ National Institute of Education, 1 Nanyang Walk, Singapore   637616 \\
$^{5}$ Institute of Advanced Studies, Nanyang Technological University, \\ 60 Nanyang View, Singapore 639673 }

\maketitle

\begin{abstract}
 We consider optimal state discrimination in a general convex operational framework, so-called generalized probabilistic theories (GPTs), and present a general method of optimal discrimination by applying the complementarity problem from convex optimization. The method exploits the convex geometry of states but not other detailed conditions or relations of states and effects. We also show that properties in optimal quantum state discrimination are shared in GPTs in general: i) no measurement sometimes gives optimal discrimination, and ii) optimal measurement is not unique.\end{abstract}

\section{Introduction}

Suppose that there is a party, say Alice, who prepares her system in a particular state. The state is chosen from a set of states that have been publicly declared. The system is then given to the other party, called Bob, who then applies a measurement to find which state has been prepared in among the possibilities. This scenario defines the problem of optimal state discrimination that seeks the guessing probability, i.e. the maximum probability that Bob can correctly guess the state that has been prepared by Alice, as well as the optimal measurement that achieves the guessing probability. Optimal state discrimination shows that there is a fundamental limit in the distinguishability of systems. This problem constitutes one of the most fundamental measures in information theory with deep connection to applications in quantum information processing \cite{ref:helstrom} \cite{ref:holevo} \cite{ref:yuen}.


Generalized probabilistic theories (GPTs) capture the formalism of the convex operational framework, in which operational significances of states, effects, and dynamics can be identified and characterized, respectively \cite{ref:birvon} \cite{ref:barnum1} \cite{ref:barrett}, see also a recent review \cite{ref:gpt}. States are elements of a convex set, effects are postulated to map states into probabilities and present probabilities measures, and dynamics constrains possible evolution of states. In quantum theory, states correspond to non-negative and unit-trace bounded operators on Hilbert spaces, effects are postulated such that product of positive-operator-valued-measures and states results to probabilities, and dynamics is generally described by positive and completely positive maps. GPTs are of fundamental interest, particularly within the foundations of quantum information theory and they also useful for identifying specific properties of states or effects that have operational significances. For instance, in quantum theory, the fact that quantum states cannot be perfectly cloned may be found as one of properties associated with the Hilbert spaces, e.g. non-orthogonality of state vectors. However, the no-cloning theorem does not necessarily rely on the structure of Hilbert spaces and in fact, GPTs which have violations of Bell inequalities can also incorporate the no-cloning theorem \cite{ref:masanes}.

Recently, optimal state discrimination in GPTs has been considered and it has been shown that it is tightly connected to ensemble steering of states and the no-signaling principle \cite{ref:jbaeep}. Specifically, in a GPT where ensemble steering is possible, the no-signaling principle can determine optimal state discrimination. This also holds true in quantum theory, where the no-signaling principle elucidates the relation between optimal state discrimination and quantum cloning \cite{ref:bae}. Given that ensemble steering itself does not single out quantum theory \cite{ref:barnum2}, the result is valid even beyond quantum theory as long as ensemble steering is allowed in a theory. That is, GPTs are a useful theoretical tool to find operational relations that may play a key role in quantum information applications \cite{ref:gpt}.

In this work, we investigate general properties of optimal state discrimination in GPTs and present a method of optimal state discrimination based on the convex geometry of a state space. After briefly introducing the framework of GPTs and optimal state discrimination, we formalize optimal state discrimination within the convex optimization framework. We show that primal and dual problems return the idential result, and thus formulate the problem in the form of the complementarity problem that gives a generalization of the optimization problems. This then allows us to derive a geometric method of state discrimination. We consider an example of GPTs, the polygon states, and apply the geometric formulation to optimal discrimination. We identify those properties that optimal quantum state discrimination shares with GPTs: i) optimal measurement is not unique in general, and ii) no measurement can sometimes give optimal state discrimination. 

The present paper is structured as follows. We first review the framework of GPTs and optimal state discrimination, and then formulate optimal state discrimination in the convex optimization framework. We show that primal and dual problems result in the same solutions, due to the strong duality in the problem. We then apply the complementarity problem that generalizes the primal and the dual problems, and derive the method of optimal state discrimination. The polygon system is considered as examples of GPTs, and we apply the method to optimal discrimination of polygon states.



\section{Optimal state discrimination in GPTs}

We briefly summarise GPTs \cite{ref:birvon} \cite{ref:barnum1} \cite{ref:barrett} and formulate optimal state discrimination as a convex optimization problem. In particular, we apply the complementarity problem and then present a method of optimal discrimination based on the convex geometry of states. 


\subsection{Generalized Probabilistic Theories}

As it has been mentioned, a GPT contains states and effects such that they produce probabilities. Any convex set can be a state space. A set of states, denoted by $\Omega$, consists of all possible states that a system can be prepared in. Any probabilistic mixture of states, i.e. $pw_1 + (1-p) w_2 \in \Omega$ for $w_{1},w_{2}\in \Omega$ and probability $p$ is also a state, and thus the set is convex. A general mapping from states to probabilities is described by {\it effects}, linear functionals $\Omega\rightarrow[0,1]$. A measurement denoted by $s$ is described by a set of effects, $E^{(s)} = \{e_{\x}^{(s)} \}_{\x=1}^{N}$, with which the probability of getting outcome $\x$ for measurement $s$ when state $w$ is given is given by $p(\x|s) = e_{\x}^{(s)}[w]$. A unit effect $u$ is introduced so that states are mapped to probabilities by effects: once a measurement occurs, we have $u[w]=1$ for all $w\in \Omega$. Thus, it holds that for any measurement $s$, we have $\sum_{\x} e_{\x}^{(s)} = u$. As effects are dual to the state space, they are also convex.

\subsection{State discrimination in convex optimization}

Optimal state discrimination in GPTs can be described by a game of two parties, Alice and Bob, as follows. Suppose that they have agreed on a set of $N$ states in advance, and then Alice prepares a system in one of the $N$ states with some probability and gives it to Bob. Note also that the {\it a priori} probabilities are known publicly. Given that the set of states and {\it a priori } probabilities are known, Bob applies measurement and attempts to guess which one has been prepared by Alice. If he makes a correct guess, the score is given $1$, and $0$ otherwise. The goal is to maximize the average score by optimizing measurements. 

Let us label the $N$ states by $\{ w_{\x}\}_{\x=1}^{N}$ and their prior probabilities by $\{ q_{\x}\}_{\x=1}^{N}$, so that together they can be expressed as $\{q_{\x}, w_{\x} \}_{\x=1}^{N}$. Bob seeks optimal measurement $\{e_{\x} \}_{\x=1}^{N}$ that fulfills the condition $\sum_{\x}e_{\x}=u$, in such a way that he makes guesses for each effect $e_{\x}$. Let $p_{B|A} (\x | \y) = e_{\x} [w_{\y}]$ denotes the probability that Bob makes a guess $w_{\x}$ from effect $e_{\x} $ corresponding to the state $w_{\y}$  given by Alice. Optimal state discrimination allows us to determine the \emph{guessing probability}, the maximum success probability that Bob makes correct guesses on average, with
\bea
p_{\g} :=  \max  \sum_{\x=1}^{N}  q_{\x} p_{B|A}(\x | \x)=   \max   \sum_{\x=1}^{N} q_{\x} e_{\x} [w_{\x}] \label{eq:gp}
\eea
where the maximization runs over all effects. Note that GPTs are generally not self-dual, that is, two spaces of states and effects are in general not isomorphic \cite{ref:brunner}.

\subsubsection{A convex optimization framework}
\label{sec:sdp}

We recall that the state space $\Omega$ is convex, and so is its dual, the space of effects, leading naturally to the following optimization problem:
\bea 
\max &~~& \sum_{\x = 1}^{N} q_{\x} e_{\x}  [ w_{\x} ] \nonumber\\
\mathrm{subject~to}&~~& ~ e_{\x} \geq 0~~\forall \x  \nonumber\\
&& ~ \sum_{\x} e_{\x} = u, \nonumber
\eea
where by $e_{\x}\geq0$ it is meant that $e_{\x} [w]\geq 0$ for all $w\in\Omega$. Note that the above problem is feasible as the set of parameters satisfying constraints is not empty. The trivial solution can be $e_{\x} =u$ for a single $\x$ and $e_{\y} =0$ for $\y \neq \x$. For convenience, we follow the notation in Ref. \cite{ref:boyd}, and rewrite the maximization problem in the above as minimization,
\bea 
\min &~~& f(\{ e_{\x} \}_{\x =1}^{N} ) =  - \sum_{\x =1}^{N} q_{\x} e_{\x} [ w_{\x} ] \nonumber\\
\mathrm{subject~to}&~~& ~ e_{\x} \geq 0~~\forall \x \nonumber\\
&& ~ \sum_{\x } e_{\x} = u, \nonumber
\eea
It is then straightforward to derive the dual problem to this. Let us write down the Lagrangian as follows,
\bea
\mathcal{L}(\{e_{\x}\}_{\x =1}^{N}, \{r_{\x}, d_{\x} \}_{\x =1}^{N}, K ) & = &  f(\{ e_{\x} \}_{\x =1}^{N} ) - \sum_{\x }  r_{\x} e_{\x} [ d_{\x} ] + (\sum_{\x} e_{\x} -u) [K], \nonumber \\
& = & - \sum_{\x} e_{\x} [q_{\x} w_{\x} + r_{\x } d_{\x} - K ] - u [ K ]. \label{eq:lag}
\eea
where $\{ r_{\x}, d_{\x}\}_{i=1}^{N}$ and $K$ are dual parameters. Note that $\{ r_{\x}\}_{\x=1}^N$ are constants and $\{d_{\x}\}_{\x=1}^N$ are normalized states. The dual problem can be obtained by solving the following,
\bea
g(\{ r_{\x}, d_{\x}\}_{\x =1}^{N},K) = \min_{\{ e_{\x} \}_{\x=1}^N} \mathcal{L} (\{e_{\x}\}_{\x=1}^{N}, \{r_{\x}, d_{\x} \}_{\x=1}^{N},K ). \label{ref:dual1}
\eea
The minimization in the above is given as $-u[K]$ if $K= q_{\x} w_{\x} + r_{\x} d_{\x}$ for each $\x$, and $-\infty$ otherwise. Thus, we have $r_{x} d_{\x} = K - q_{\x} w_{\x}$ for each $\x$. Since $r_{\x} d_{\x}$ is a (unnormalized) state, it is positive, that is $e [ d_{\x} ] \geq0$ for all effects $e$. We write this as, $K\geq q_{\x} w_{\x}$ for each $\x$. The dual problem is thus as follows,
\bea
\max&~~& -u [ K ] \nonumber\\
\mathrm{subject~to} &~~& K\geq q_{\x} w_{\x},~~\forall \x. \nonumber 
\eea
Or, equivalently,  
\bea
\min&~~& u [ K ] \nonumber\\
\mathrm{subject~to} &~~& K \geq q_{\x} w_{\x},~~\forall \x.  \nonumber 
\eea
In the above, the inequality means an order relation in a convex cone, which is determined by effects, that is, $e [ K-q_i w_i ] \geq 0 $ for all effects $e$. Note also that the dual problem is also feasible: the trivial solution would be $K= \sum_{\x} q_{\x} w_{\x}$.

\subsection{Constraint Qualification}

Recall that in general, the primal and the dual problems do not return an identical solution but there can be a finite gap between solutions of the two problems. In the case of state discrimination in the above, both problems in the above are feasible. This means that from Slater's constraint qualification, the strong duality holds. Hence, no gap exists between the solutions, and in other words, one can get the optimal solution by solving either of the primal or the dual problems.

In addition, the strong duality also implies that the list of optimality conditions, the so-called Karush-Kuhn-Tucker (KKT) conditions, are also sufficient. That is, parameters satisfying KKT conditions provide optimal solutions in both primal and dual problems. For the optimization problems in the above, The KKT conditions are, together with constraints in both primal and dual problems, as follows,
\bea
K & = &  q_{\x} w_{\x} + r_{\x}d_{\x} ,~~\forall ~ \x \nonumber  \\
e_{\x } [ r_{\x}d_{\x} ] & = & 0, ~~\forall~\x. \nonumber
\eea
The former one is called Lagrangian stability, and the latter one the complementary slackness. The fact that the strong duality holds also guarantees that there exist dual parameters $K$ and $\{ r_{\x}, d_{\x} \}_{|x=1}^{N}$ that fulfill KKT conditions, and then those parameters give optimal solutions to the primal and the dual problems. Here, the optimal effects are also characterized by the complementary slackness in the above. This also shows existence of optimal effects or observables in a GPT. All these follow from the fact that the state space is convex. For comparison with quantum cases, the formulation for minimum-error discrimination has been shown in Ref. \cite{ref:baegeo}, and see also its applications to various forms of figures of merit in Ref. \cite{ref:kato}.

To summarize, the sole fact that state and effect spaces are convex allows us to formalize the discrimination problem in the convex optimization framework \cite{ref:boyd}. This in fact provides a general approach of finding optimal discrimination in GPTs. For states $\{q_{\x},w_{\x} \}_{\x=1}^{N}$, we take the form in Eq. (\ref{eq:gp}) as the primal problem denoted by $p^{*}$ and derive its dual $d^{*}$, as follows,
\bea
\mathrm{Primal: }~~p^{*} & = & \max\{ \sum_{\x=1}^{N} q_{\x} e_{\x} [ w_{\x} ] ~~|~~ e_{\x}\geq0 ~\forall\x, ~\sum_{\x=1}^{N} e_{\x} = u  \}   \label{eq:primal} \\
\mathrm{Dual: }~~ d^{*} & = & \min \{ u [K ]~ ~|~ ~K\geq q_{\x} w_{\x},~\x=1,\cdots,N \} \label{eq:dual}
\eea
where inequalities mean the order relation in the convex set: by $e_{\x}\geq 0$, it is meant that $e_{\x}[w]>0$ for all $w\in \Omega$, and by $K\geq q_{\x} w_{\w}$, that $e[K-q_{\x} w_{\x} ]\geq 0$ for all effects $e$.


For the primal and dual problems in Eqs. (\ref{eq:primal}) and (\ref{eq:dual}), the property called the strong duality holds true. This means that the two problems have an identical solution, i.e. $p^{*} = d^{*}$, and therefore one can obtain the guessing problem by solving either of the problems. The strong duality can be shown from the so-called Slater's constraint quantification in convex optimization. A sufficient condition for the strong duality is the strict feasibility to either of primal or dual problems, that is, the existence of a strictly feasible point of parameters. For instance, primal parameters $\{e_{\x} = u/N\}_{\x=1}^{N}$ are in the case, since $e_{\x} [w_{\y}]>0$ $\forall \x, \y$ and $\sum_{\x} e_{\x}=u$. From this, it is shown that the guessing probability can be obtained from either the primal or the dual problem.

\subsection{ The complementarity problem}

In convex optimization, there is another approach called \emph{the complementarity problem} that generalizes primal and dual problems. It collects optimality conditions and analyzes them directly. Consequently, the complementarity problem deals with both primal and dual parameters in Eqs. (\ref{eq:primal}) and (\ref{eq:dual}) and find all of optimal parameters. In this sense, the approach is generally not considered more efficient than primal or dual problems. The advantage, actually, lies at the fact that generic structures existing in an optimization problem are found and exploited. 

The optimality conditions for optimal state discrimination in Eqs. (\ref{eq:primal}) and (\ref{eq:dual}) can be summarized by the so-called Karush-Kuhn-Tucker (KKT) conditions, which are constraints listed in Eqs. (\ref{eq:primal}) and (\ref{eq:dual}), together with the followings,
\bea
\mathrm{ (Symmetry~parameter) }&&~ K = q_{\x} w_{\x} + r_{\x} d_{\x} ,~\forall ~\x \label{eq:kkt1} \\
\mathrm{(Orthogonality)}&&~ e_{\x} [r_{\x} d_{\x}]  = 0, ~\forall~\x, \label{eq:kkt2}
\eea
where $r_{\x} \in [0,1] $ for all $\x$, and $\{ d_{\x}\}_{\x=1}^{N}$ are normalized states, i.e. $u [ d_{\x} ]=1$. We call $\{ d_{\x}\}_{\x=1}^N$  complementary states that construct the symmetry operator. Two conditions in the above are explained in terms of the convex geometry of given states, as follows. 
\begin{enumerate}
\item The first condition, symmetry parameter, follows from the Lagrangian stability and shows that for any discrimination problem e.g. $\{ q_{\x}, w_{\x} \}_{\x=1}^{N}$, there exists a single parameter $K$ which is decomposed into $N$ different ways with given states and complementary states $\{r_{\x}, d_{\x}\}_{\x=1}^{N}$. Then, the second condition in Eq. (\ref{eq:kkt2}) from the complementary slackness characterizes optimal effects by the orthogonality relation between complementary states and optimal effects. These generalize optimality conditions from quantum cases to all GPTs, see also various forms of optimality conditions in quantum cases \cite{ref:baegeo}.  

\item Primal and dual parameters satisfying the KKT conditions are automatically optimal parameters that provide solutions in the primal and the dual problems. Note also that, since the strong duality holds, both problems show the same solution. Conversely, the fact that the strong duality holds in Eqs. (\ref{eq:primal}) and (\ref{eq:dual}) implies the existence of optimal parameters which satisfy KKT conditions and give the guessing probability in Eq. (\ref{eq:gp}). 
\end{enumerate}

Note that a similar approach has been made in Ref. \cite{ref:kimuragpt} in the form of the so-called Helstrom family, by generalising examples in quantum cases to GPTs. For quantum state discrimination, the approach based on the complementarity problem has been firstly applied in Refs. \cite{ref:Hwang} \cite{ref:HwangBae} for two qubit states. This has been generalised to a pair of arbitrary states in GPTs \cite{ref:kimura2}. When this result is generalized to arbitrary number of states in GPTs, however, the existence of the symmetry operator and the orhogonality conditions has been only assumed \cite{ref:kimuragpt}: in particular, those cases for which the optimal parameters exist are called Helstrom families. Here, we apply the complementarity problem that immediately proves the existence of optimal parameters.  

\subsection{ The geometric method and the general form of the guessing probability}

We are now ready to present a geometric method of solving minimum-error state discrimination within the framework of GPTs for the complementarity problem. We first observe that, for optimality conditions in Eqs. (\ref{eq:kkt1}) and (\ref{eq:kkt2}), constraints for states and effects are separated. The symmetry parameter $K$ is characterized on a state space and gives the guessing probability, see Eq. (\ref{eq:dual}), that is, 
\bea
p_{\g} = u [K ] = q_{\x} + r_{\x}. \label{eq:g}
\eea  
This means that one can find the guessing probability from a state space. To do this, one has to find the symmetry operator $K$ such that it is decomposed into a given state $q_{\x} w_{\x}$ and complementary states $r_{\x} d_{x}$ in the state space. Or, equivalently, one has to search complementary states $\{r_{\x},d_{\x} \}_{\x=1}^{N}$ fulfilling Eq. (\ref{eq:kkt1}) on the state space. 

Let us introduce a convex polytope denoted by $\mathcal{P}(\{ q_{\x}, w_{\x} \}_{\x=1}^{N})$ of given states in the state space: each vertex of the polytope corresponds to state $q_{\x} w_{\x}$ for $\x=1,\cdots,N$. Then, the polytope of complementary states, $\mathcal{P}(\{ r_{\x}, d_{\x} \}_{\x=1}^{N})$, is immediately congruent to $\mathcal{P}(\{ q_{\x}, w_{\x} \}_{\x=1}^{N})$ in the state space: from Eq. (\ref{eq:kkt1}) the following holds,
\bea
q_{\x} w_{\x} - q_{\y} w_{\y} = r_{\y} d_{\y} - r_{\x} d_{\x},~~\mathrm{for~all}~\x,\y, \label{eq:kkt}
\eea
which shows that corresponding lines of two polytopes $\mathcal{P}(\{ q_{\x}, w_{\x} \}_{x=1}^{N})$ and $\mathcal{P}(\{ r_{\x},d_{\x} \}_{x=1}^{N})$ are of equal lengths and anti-parallel. Then, from the convex geometry of the state space, one can find the polytope of complementary states as well as complementary states by putting two congruent polytopes such that the condition in Eq. (\ref{eq:kkt1}) holds. Once complementary states are obtained, optimal effects can be found from the orthogonal relation in Eq. (\ref{eq:kkt2}), accordingly. 

Finally, let us provide a general form of the guessing probability in GPTs, when {\it a priori} probabilities are equal i.e. $q_{\x} = 1/N$ for all $\x$. In this case, the guessing probability is in a simpler form and show its meaning with the convex geometry. First, from Eq. (\ref{eq:g}) we have $p_{\g} = q_{\x} +r_{\x}$ for any $\x$. Since $q_{\x}=1/N$, we have $r_{\x} = r_{\y}$ for all $\x,\y$. Denoted by $r:=r_{\x}$ for all $\x$, the guessing probability has the form in the following
\bea 
p_{\g} = \frac{1}{N} + r,~~\mathrm{with}~~r = \frac{ \| \frac{1}{N} w_{\x} - \frac{1}{N} w_{\y}  \| }{ \|  d_{\x} - d_{\y}  \|}\label{eq:r}
\eea
where the expression of $r$ follows from the condition in Eq. (\ref{eq:kkt}) with a distance measure $\| \cdot\|$ that can be defined in the state space. The parameter $r$ has a meaning as the ratio between two polytopes, $\mathcal{P}(\{ 1/N, w_{\x} \}_{x=1}^{N})$ of given states, and  $\mathcal{P}(\{d_{\x} \}_{x=1}^{N})$ of complementary states.




\section{ Examples: polygon states}

We illustrate the method of optimal state discrimination in GPTs, with an example called the polygon systems in Ref. \cite{ref:brunner}. We consider cases of three and four states and apply the geometric method of optimal state discrimination. It is straightforward to apply to $N$ states. The polygon system is in general given by $N$ states $\{w_{\x} \}_{\x=0}^{N-1}$, 
\bea
w_{\x} = \left( \begin{array}{ccc}
r_n  \cos \frac{ 2\pi \x  }{ n} \\
r_n  \sin \frac{ 2\pi \x  }{ n} \\
1
\end{array} \right)  \nonumber
\eea
where $r_n = \cos^{-1/2} (\pi/n)$. Effects are given by $\{ f_{\x}\}_{\x =0}^{N-1} $ as follows,
\bea
\mathrm{for~ even~ N}, ~~ f_{\x}  = \frac{1}{2} \left( \begin{array}{ccc}
r_n \cos \frac{(2\x-1) \pi}{n}  \\
r_n \sin \frac{(2\x-1) \pi}{n}   \\
1
\end{array} \right),~ ~~\mathrm{and ~for~ odd~ N}, ~~  f_{\x}  = \frac{1}{1+r_{n}^2} \left( \begin{array}{ccc}
r_n \cos \frac{ 2 \pi \x  }{n}    \\
r_n \sin \frac{ 2\pi \x   }{n}     \\
1
\end{array} \right)
\label{eq:ex}
\eea
where the unit effect $u=(0,0,1)^{T}$ and the map to probabilities is given by the Euclidean inner product between states and effects. 


\begin{figure}
\begin{center}  
\includegraphics[width= 10cm]{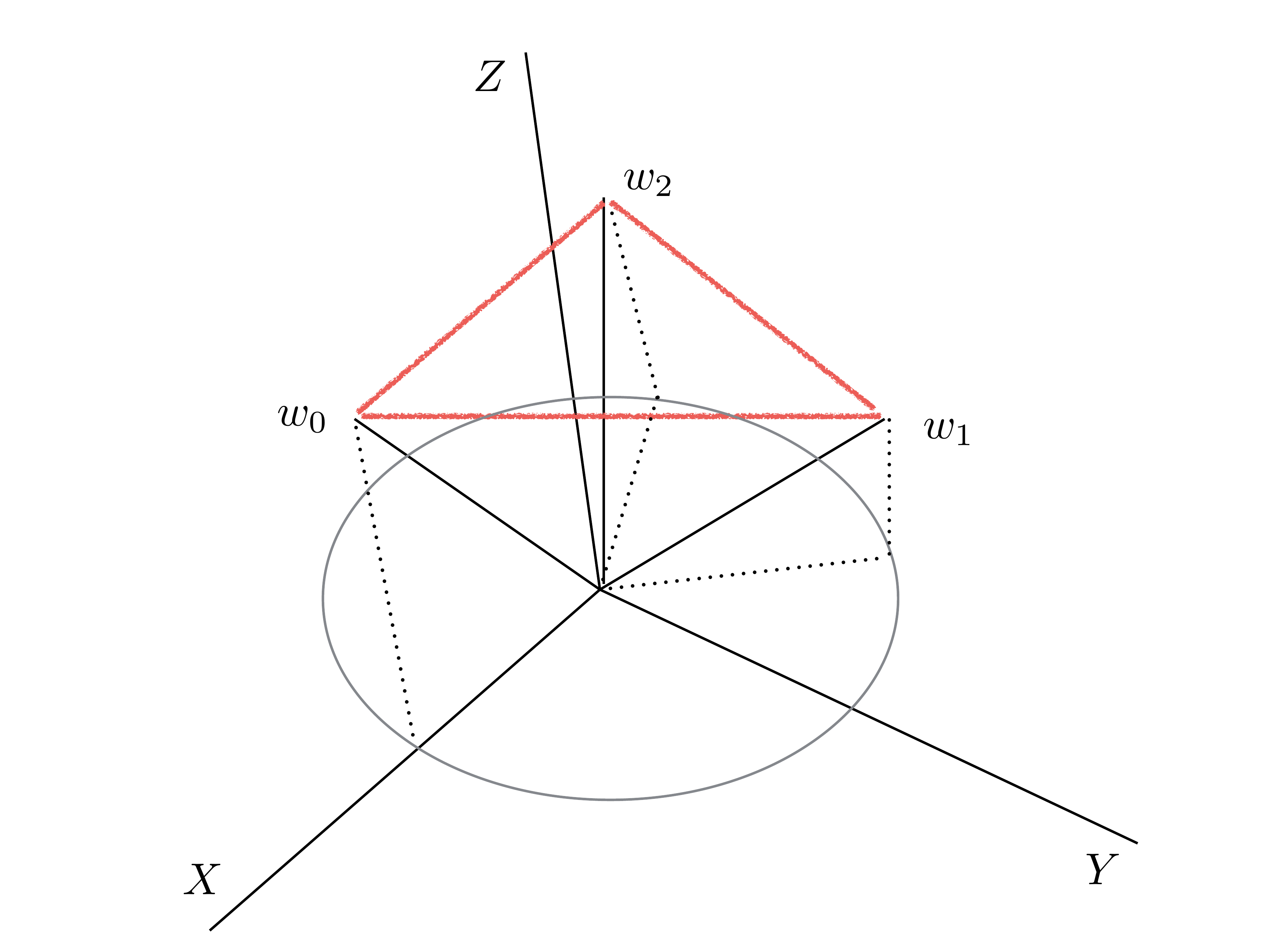}
\caption{The polygon states for $N=3$ are shown, see also Eq. (\ref{eq:3state}). The three states form a regular triangle at the place $z=1$. The effects in Eq. (\ref{eq:3effects}) are identical to the states.}\label{fig:3}
\end{center}
\end{figure}

\subsection{A case of $N=3$} 

Let us first consider the case $N=3$, in which states and effects are given as 
\bea
w_{0} = \left( \begin{array}{ccc}
\sqrt{2} \\
0 \\ 
1
\end{array} \right),~  
w_{1} = \left( \begin{array}{ccc}
-\sqrt{2}/2 \\
 \sqrt{6} /2  \\ 
1
\end{array} \right),~  
w_{2} = \left( \begin{array}{ccc}
- \sqrt{2} / 2  \\
- \sqrt{6}/ 2  \\ 
1
\end{array} \right),~  
 \label{eq:3state}
\eea
\bea
f_{0} = \frac{1}{3} \left( \begin{array}{ccc}
\sqrt{2} \\
0 \\ 
1
\end{array} \right),~  
f_{1} = \frac{1}{3}\left( \begin{array}{ccc}
-\sqrt{2}/2 \\
 \sqrt{6} /2  \\ 
1
\end{array} \right),~  
f_{2} = \frac{1}{3} \left( \begin{array}{ccc}
- \sqrt{2} / 2  \\
- \sqrt{6}/ 2  \\ 
1
\end{array} \right).~  
 \label{eq:3effects}
\eea
One can easily check that $f_0+ f_1 + f_2 =u$. We consider optimal state discrimination for $\{1/3, w_{\x}\}_{\x=0}^2$. Applying the geometric method and also from the geometry of the polygon system for $N=3$, see also Fig. (\ref{fig:3}) one can find it holds that
\bea
K = \frac{1}{3} w_{\x} + \frac{2}{3} d_{\x},~~\mathrm{with}~ d_{\x} = \frac{1}{2} (f_{\x+1} + f_{\x+2}), \nonumber
\eea
where $\x=0,1,2$ and the addition is computed in modulo $3$. Optimal measurement is therefore $\{f_{\x} \}_{\x=0}^2 $ since $f_{\x} [d_{\x}] =0$, from the orthogonality condition in Eq. (\ref{eq:kkt2}). Thus, we have $p(\x|\x) = f_{\x} [w_{\x}]  = 1$ for $\x=0,1,2$, and the guessing probability is found as 
\bea
p_{\g} = \frac{1}{3} \sum_{\x=0}^2 p(\x | \x) =1, \nonumber
\eea
which leads to the perfect discrimination. 

\begin{figure}
\begin{center}  
\includegraphics[width= 10cm]{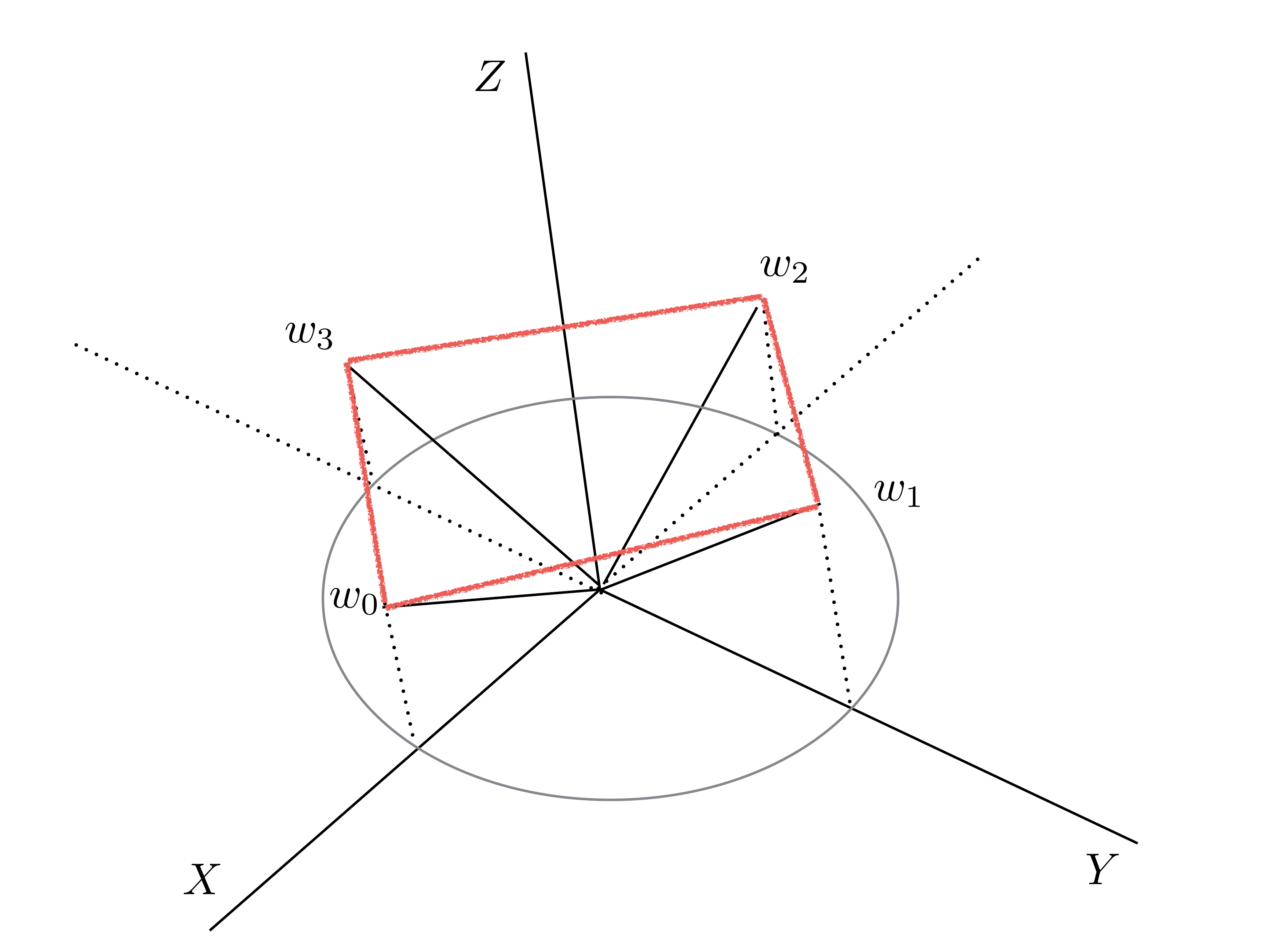}
\caption{The polygon states for $N=4$ are shown, see also Eq. (\ref{eq:4state}). The four states form a square at the place $z=1$. The effects in Eq. (\ref{eq:4effects}) are located at the place $z=1/2$. }\label{fig:4}
\end{center}
\end{figure}

\subsection{A case of $N=4$}

We next consider the case $N=4$, in which states and effects are given as 
\bea
w_{0} = \left( \begin{array}{ccc}
r_4 \\
0 \\ 
1
\end{array} \right),~  
w_{1} = \left( \begin{array}{ccc}
 0  \\
r_4  \\ 
1
\end{array} \right),~  
w_{2} = \left( \begin{array}{ccc}
- r_4  \\
0 \\ 
1
\end{array} \right),~  
w_{3} = \left( \begin{array}{ccc}
0  \\
- r_4 \\ 
1
\end{array} \right),~  
 \label{eq:4state}
\eea
\bea
f_{0} = \frac{1}{2} \left( \begin{array}{ccc}
 r_4 / \sqrt{2}  \\
- r_4 / \sqrt{2}  \\ 
1
\end{array} \right),~  
f_{1} = \frac{1}{2}\left( \begin{array}{ccc}
 r_4 / \sqrt{2}  \\
 r_4 / \sqrt{2}   \\ 
1
\end{array} \right),~  
f_{2} = \frac{1}{2} \left( \begin{array}{ccc}
-  r_4 / \sqrt{2}  \\
 r_4 / \sqrt{2}  \\ 
1
\end{array} \right),~  
f_{3} = \frac{1}{2} \left( \begin{array}{ccc}
- r_4 / \sqrt{2 } \\
- r_4 / \sqrt{2}  \\ 
1
\end{array} \right).~  
 \label{eq:4effects}
\eea
For four states $\{ 1/4 ,w_{\x} \}_{\x=0}^{3}$, the goal is now to find the guessing probability and optimal measurement. Exploiting the convex geometry, see Fig. (\ref{fig:4}), the polytope $\mathcal{P}(\{ 1/4, w_{\x} \}_{x=0}^{3})$ forms a rectangle, from which it follows that $r=1/4$ from Eq. (\ref{eq:r}). To be precise, from the state space geometry, one can see that
\bea
K = \frac{1}{4} w_{\x} + \frac{1}{4} d_{\x},~\mathrm{where}~ d_{\x} = w_{\x+2},~\mathrm{for}~\x=0,1,2,3,~ ~(\mathrm{mod}~4), \nonumber
\eea
where the complementary states are obtained as $d_{\x} = w_{\x+2}$. Thus, we have the guessing probability, from the primal problem in Eq. (\ref{eq:kkt1}),
\bea
p_{\g} = u[K] = \frac{1}{2}. \nonumber
\eea  
Note that these four states are analogous to cases in quantum theory, pairs of orthogonal states: for the four quantum states, the guessing probability is also given by $1/2$ \cite{ref:baegeo}. 


Optimal measurement is obtained by using the orthogonality condition in Eq. (\ref{eq:kkt2}). In fact, optimal measurement is not unique and the following effects give the guessing probability.  
\begin{itemize}
\item  i) $\{ f_{\x}/2 \}_{\x=0}^{3}$: In this case, we have 
\bea
p(\x | \x ) = \frac{1}{2} f_{\x} [w_{\x}] = \frac{1}{2},~\mathrm{and~thus,~} p_{\g} = \frac{1}{4} \sum_{x=0}^3 p(\x | \x) = \frac{1}{2}.\nonumber
\eea
One can also easily check the orthogonality condition $(1/2)e_{\x}[d_{\x}] =0$ and also that $\sum_{\x} e_{\x} /2 =u$.

\item  ii) $\{f_{0}, f_{2} \}$: In this case, measurement on effect on $e_{0}$ concludes that given state is either $w_0$ or $w_3$, and $e_{3}$ to  $w_1$ or $w_2$. This is because, from the orhogonality condition in Eq. (\ref{eq:kkt2}), it holds that
\bea
f_{0} [d_0] = f_0[w_2 ]=0, ~\mathrm{and}~  f_{0} [d_3] =e_0[w_1] = 0. \nonumber
\eea
Once measurement on effect $f_{0}$ ($f_2$) is found, one randomly conclude the given state is either $w_0$ or $w_3$ ($w_1$ or $w_2$), and the guessing probability is obtained $1/2$. 

\item  iii) $\{ f_{1}, f_{3} \}$: This case works in a similar way to the previous. Measurement on effect on $f_{1}$ concludes that given state is either $w_0$ or $w_1$, and $f_{3}$ to  $w_2$ or $w_3$.
\end{itemize}

From optimal measurement shown in the above, we remark that properties of optimal quantum state discrimination also hold true in GPTs. First, optimal measurement of quantum state discrimination is generally not unique \cite{ref:helstrom}, and the example in the above shows that this also holds true in GPTs. Moreover, optimal measurement for discriminating $N$ quantum states does not always contain output ports in the same number, that is, $N$ POVMs \cite{ref:baegeo} \cite{ref:hunter}. This also holds true in GPTs in general as shown above.

\subsection{When no measurement is optimal} 

We here show another property that quantum state discrimination shares with GPTs. Namely, no measurement is optimal in state discrimination. That is, applying no measurement but simply guessing the state from {\it a priori} probabilities gives a guessing probability higher than any other strategies. This also holds true in GPTs. In the following, we provide an example from the result in the quantum case \cite{ref:hunter}.

Let us consider the four polygon states $\{ w_{\x} \}_{\x=0}^3$ for $N=4$ in the above together with their mixture $w_4 = \sum_{\x =0}^3 w_{\x} / 4$. Let $q_{\x} = (1-p)/4$ denote {\it a priori} probabilities for states $w_{\x}$ for $\x=0,1,2,3$, respectively, and $q_4=p$ for state $w_4$. Hence, we consider optimal state discrimination for $\{ q_{\x} , w_{\x} \}_{\x=0}^4$.  In particular, let us also assume that $p\geq 1/5$. Then, one can find the optimal discrimination with the symmetry operator as follows,
\bea
K & = &  p~ w_4  \nonumber \\
& = & \frac{1-p}{4} w_{\x} + r_{\x} d_{\x},~\mathrm{for}~\x = 0,1,2,3 \label{eq:decom}
\eea
with constants $ \{ r_{\x} = ( 1+5p) / 4 \}_{\x=0}^{3}$. It is then straightforward to find $\{ d_{\x} \}_{\x=0}^{3} $ such that the equalities in Eq. (\ref{eq:decom}) hold true. Note also that whenever $p\geq 1/5$ it holds that $ r_{\x} d_{\x} \geq 0$. Then, the guessing probability is simply given as $p_{\g} = u(K) = p$, which can be made by guessing the state $w_4$ with the {\it a priori} probability, without measurement.


\section{Conclusions}

Optimal state discrimination is one of the most fundamental tasks in information theory, and also connected to information applications. For instance, in quantum cases it is the operational task that corresponds to the information-theoretic measure, the min-entropy \cite{ref:renner}. On the other hand, GPTs are of theoretical and fundamental interest such that states, effects, and dynamics are identified in a convex operational framework. Their operational significances can be found without detailed structures of a given theory, e.g. Hilbert spaces of quantum theory. 

In the present work, we have considered optimal state discrimination in GPTs within the convex optimization framework. This generalizes the result in quantum cases where optimization runs over symmetric operators describing quantum states and measurements \cite{ref:baegeo}. Here, we have considered optimal state discrimination without such structures, and shown that the results in quantum cases, e.g. see Ref. \cite{ref:baegeo}, are shared in GPTs in general. These include, firstly, the convex optimization and the complementariy problem, and then the method of optimal state discrimination with the convex geometry of state spaces. In particular, we has shown with the polygon systems how the method can be applied. We have shown that the followings hold true in GPTs in general: i) optimal measurement is not unique, and ii) no measurement can sometimes give optimal discrimination. The results may be useful in the operational characterization of quantum information processing, and we also envisage their usefulness in quantum information applications.

\section*{Acknowledgement}

This work is supported by the research fund of Hanyang University (HY-2015-259), People Programme (Marie Curie Actions) of the European Union's Seventh Framework Programme (FP7/2007-2013) under REA grant agreement N. 609305, and the National Research Foundation \& Ministry of Education, Singapore. 
 





\end{document}